\documentclass[twocolumn,aps,superscriptaddress]{revtex4-1}
\usepackage{palatino}
\usepackage[latin1]{inputenc}
\usepackage{epsf}
\usepackage{amsmath,amssymb}
\usepackage{latexsym}
\usepackage{calc}
\usepackage{color}
\usepackage{shadow}
\usepackage{epsfig}
\usepackage{braket}

\newcommand{\ben}{\begin{equation}}
\newcommand{\een}{\end{equation}}
\newcommand{\bea}{\begin{eqnarray}}
\newcommand{\eea}{\end{eqnarray}}

\def\sss{\scriptscriptstyle\rm}
\def\x{_{\sss X}}
\def\c{_{\sss C}}
\def\s{_{\sss S}}
\def\xc{_{\sss XC}}

\def\H{_{\sss H}}

\def\ext{_{\rm ext}}

\def\br{{\bf r}}

\def\p{{\partial}}

\begin{document}
\title{Absence of Dynamical Steps in the Exact Correlation Potential in Linear Response}
\author{Kai Luo}
\address{Department of Physics and Astronomy, Hunter College and the Graduate Center of the City University of New York, 695 Park Avenue, New York, New York 10065, USA}
\author{Peter Elliott}
\address{Max-Planck-Institut f\"{u}r Mikrostrukturphysik, Weinberg 2, 06120 
Halle (Saale), Germany}
\author{Neepa T. Maitra}
\address{Department of Physics and Astronomy, Hunter College and the Graduate Center of the City University of New York, 695 Park Avenue, New York, New York 10065, USA}

\date{\today}
\pacs{}

\begin{abstract}
Recent work ~\cite{EFRM12} showed that the exact
exchange-correlation potential of time-dependent density-functional
theory generically displays dynamical step structures. These have a
spatially non-local and time-non-local dependence on the density in
real time dynamics. The steps are missing in the usual approximations
which consequently yield inaccurate dynamics. Yet these same
approximations typically yield good linear response spectra. Here we
investigate whether the steps appear in the linear response regime,  when the response is calculated from a real-time dynamics simulation, by examining the exact correlation potential of model two-electron systems at various times. 
We find there are no step structures in regions where the system response is linear. Step structures appear in the correlation
potential only in regions of space where the density response is
non-linear; these regions, having exponentially small density, do not
contribute to the observables measured in linear response.
\end{abstract}
\maketitle 

\section{Introduction}
Time-Dependent Density Functional Theory (TDDFT) is now a method of
choice for the calculation of excitation spectra and response
properties in materials science and quantum
chemistry~\cite{RG84,newTDDFTbook,Carstenbook}. It maps the system of
interacting electrons into a fictitious one of non-interacting
fermions, called the Kohn-Sham system, from which all
properties of the original system may be exactly extracted in principle. Therefore large
systems of relevance in biochemistry and nanoscience may be treated. 
The Kohn-Sham fermions evolve in a one-body potential, $v\s(\br,t)$,
which has the property that the exact one-body density of the system
of interacting electrons is exactly reproduced by the non-interacting
Kohn-Sham fermions. However, an essential component of this potential,
the exchange-correlation (xc) potential, is unknown, and must be
approximated, as a functional of the density, the interacting initial
state $\Psi_0$ and non-interacting initial state $\Phi_0$:
$v\xc[n,\Psi_0,\Phi_0](\br,t)$. The vast majority of applications
today use an adiabatic approximation, meaning one where the
instantaneous density is input into a ground-state xc approximation:
$v\xc^{A}[n,\Psi_0,\Phi_0](\br,t) = v\xc^{\rm g.s.}[n(\br,t)]$. All
memory-dependence is neglected. The adiabatic approximation is behind
the linear response results whose success has propelled TDDFT forward,
and it is implicitly assumed in all the readily available codes today.
Cases for which the adiabatic approximation fail are known
(e.g. states of double-excitation character, long-range
charge-transfer excitations between open-shell fragments, conical intersections), and users are
generally aware to apply caution when interpreting the TDDFT results
in these cases. Still, TDDFT has proven itself remarkably useful in
its balance between accuracy and efficiency for spectra and response,
and functional developments are on-going~\cite{newTDDFTbook,Carstenbook,CH12}. 

TDDFT is not limited to the linear response regime: indeed given the
dearth of alternative practical methods of solving correlated electron
dynamics in non-equilibrium situations, the non-linear regime is
arguably more important for TDDFT. Moreover, due to the recent intense
progress in attosecond technology, the control and study of electron
dynamics, with concommitant control and study of nuclear dynamics, are
becoming an experimental reality. However, for real-time
non-equilibrium dynamics, much less is known about the accuracy of the
usual functionals in TDDFT, and from comparisons with the few
available exactly solvable systems, it appears that memory effects, missing in the usual adiabatic approximations, can be significant. 

Recent work has shown the prevalence of dynamical step structures in
the time-dependent exchange-correlation potential in
far-from-equilibrium situations~\cite{RG12,EFRM12,FERM13}. These step
structures were found to arise in a variety of dynamics, from resonant
Rabi oscillations in a model atom and molecule~\cite{EFRM12,FERM13},
to dynamics under an arbitrary strong field~\cite{EFRM12}, to
quasiparticle propagation in a semiconducting wire~\cite{RG12}. The
steps were found to have a very non-local dependence on the density in
both space and time; it was shown that even an adiabatically-exact
approximation fails to capture them. Typical approximations in use
today do not contain these structures, resulting in inaccurate
dynamics, as shown in the examples in ~\cite{EFRM12,FERM13}. Yet these
same approximations do give good spectra for these systems. The
question then arises: what happens to these steps in the linear
response regime? In this paper, we show that the steps are in fact a
{\it nonlinear} response phenomenon, and do not appear when the
response of the system is linear.  To show this, we calculate the
time-dependent correlation potential in model two-electron systems
under several linear response situations, including a weak field
smoothly turned on and off, as well as evolution under a delta-kick.
A non-adiabatic kernel has been shown to be essential to obtain
excitations of double-excitation character~\cite{MZCB04, THp}, but we show
here, that the non-adiabatic step of Ref.~\cite{EFRM12} is an
unrelated phenomenon.

The rest of the paper is organized as follows. In
Section~\ref{sec:LRdynamics}, we first introduce the model systems
used in our study. We then proceed to find the time-dependent
correlation potential in linear response to a smoothly turned on weak
field (Sec.~\ref{sec:gaussian}), and to a delta-kick
(Sec.~\ref{sec:delta}). In Sec.~\ref{sec:analysis} we find explicit
expressions for the terms in the correlation potential that scale
linearly with the system's response, and finally Sec.\ref{sec:summary}
contains our conclusions.

\section{Dynamics in the linear response regime}
\label{sec:LRdynamics}
The system we will mostly focus on in this paper is a one-dimensional  (1D) model
of the He atom; 
the Hamitonian can be written as
\ben
\hat{H} = \hat{H}_0 +\hat{H}_1(t) = \hat{T} + \hat{V}(t) + \hat{W}
\label{eqn:hamitonian},
\een
where $\hat{T} = \sum_i -\frac{1}{2} \frac{\p^2}{\p x_i^2}$ is the kinetic energy, $\hat{V}(t) = \sum_i [-2/\sqrt{x_i^2+1} -x_i {\cal E}(t)]$ is the external potential, and
$\hat{W} = 1/\sqrt{(x_1-x_2)^2 + 1}$ is the soft-Coulomb electron-electron interaction~\cite{JES88}.  The sums go over two fermions. (Atomic units $e^2 = \hbar=m_e=1$) are used throughout the paper). 
This soft-Coulomb model is commonly used in analyzing
functionals, since it is numerically straightforward to find the exact time-evolving wavefunction, and then extract the
exact exchange and correlation potentials for comparison with
approximations~\cite{VIC96,BN02,KLEG01,LL98,WB06,TGK08,TK09,TMM09,FHTR11}.
We will apply weak off-resonant fields, represented by ${\cal E}(t)$ above,  to stimulate linear response of the system, as will be detailed below.

Since all double-excitations in the He atom lie in the continuum, we
instead consider a model of a quantum dot to study this issue,
taking $\hat{V}(t) = \sum_i[\frac{1}{2}x_i^2 + x_i^2 {\cal
  F}(t)]$. The time-dependent driving in this case is modified
  from the usual dipole form to a quadratic form, since linear dipole pertubation only couples to the first excited state which is predominantly a single excitation~\cite{EM11,D94}.

For two electrons in a spin-singlet, choosing the
initial KS state as a doubly-occupied spatial orbital, $\phi(\br,t)$,
means that the exact KS potential for a given density evolution can be
found easily (see e.g. Ref.~\cite{HMB02,EFRM12}): in 1D,
we have
\ben
\label{eqn:vsexact}
v\s(x,t)= -\frac{(\partial_x n(x,t))^2}{8n^2(x,t)} +\frac{\partial^2_x n(x,t)}{4 n(x,t)} - \frac{ u^2(x,t)}{2} - \int^x\frac{\partial u(x',t)}{\partial t}dx'
\een 
where $u(x,t)=j(x,t)/n(x,t)$ is the local ``velocity'', $n(x,t)$ is the one-body density, and $j(x,t)$ is the current-density.
We numerically solve the exact
time-dependent two-electron wavefunction,
 obtain the one-body density and
current-density, and insert them into Eq.~\ref{eqn:vsexact}. 
The exchange-potential in this case is simply minus half the Hartree potential, $v\x(x,t) =
-v\H(x,t)/2$, with $v\H(x,t) = \int w(x',x)n(x',t) dx'$, in terms of the two-particle interaction $w(x',x)$. Therefore,
we can directly extract the correlation potential using
\ben
\label{eqn:vcdef}
v\c(x,t) = v\s(x,t) - v\ext(x,t) -v\H(x,t)/2\;,
\een
where $v\ext(x,t)$ is the external potential applied to the system.

{\it Computational details}: We use {\tt
  octopus}~\cite{octopus,octopus2} to compute the exact
wavefunction. The time-dependent Schr\"odinger equation is solved by
first mapping the Hamiltonian of two interacting electrons in 1D  onto the Hamiltonian of one electron in 2D.
We use a grid of size $40.00$au and grid spacing of $0.1$au. The approximated enforced time-reversal symmetry method was used in the propagation, with a time-step of $0.001$au.
The densities and
current densities are then extracted and a
 standard finite-difference
scheme is used get the time derivative of the velocity. 

\subsection{Dynamics in a Gaussian-Shaped Pulse}
\label{sec:gaussian}
The examples in Ref.~\cite{EFRM12}  began in the ground state and 
either applied a weak resonant field or a strong arbitrary field to the system, 
or began in a superposition of a ground and excited state. 
None of these situations are the territory of linear response.
Instead here, we apply a weak off-resonant field, but with an envelope such that a number of excitations fall under it. 
To this end, we apply a weak electric field ${\cal E}(t)$ with the following Gaussian envelope:
\ben
{\cal E}(t) = \epsilon_\alpha \; e^{-\left[\frac{t-3 T_0}{\sqrt{2} T_0}\right]^2}\cos(\Omega_0 t)\;,
\label{eqn:field}
\een
where $T_0 = 2 \pi/\Omega_0$ is the period corresponding to the
central frequency, and $\epsilon_\alpha$ is the peak field strength (see below). 
Figure~\ref{fig:field} shows the power spectrum for strength $\epsilon_1$; excitations of the 1D He model of frequency $0.533, 0.672, 0.7125...$au lie in its bandwidth. Here we have chosen 
$\Omega_0 = 0.7au$, but our conclusions are independent of this value.
\begin{figure}[h]
	\begin{center}
	\includegraphics[width=0.25\textwidth,angle=270,clip]{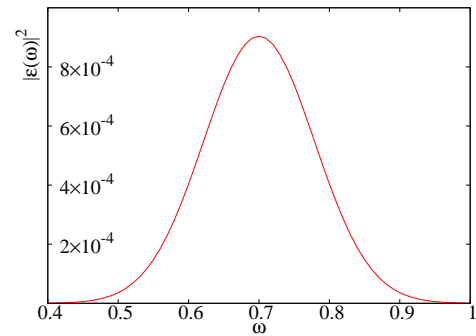} 
	\end{center}
	\caption{\label{fig:field}(color online). This plot shows the modulus square of the Fourier-transformed field of strength of $\epsilon_1 = 0.0067$au, which includes the first several excitation energies. }
\end{figure}
\begin{figure}[h]
	\begin{center}
	\includegraphics[width=0.475\textwidth,height=0.4\textwidth,clip]{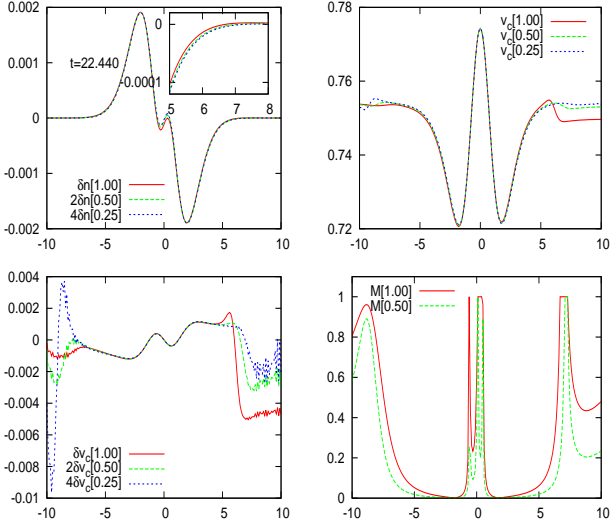} 
	\end{center}
	\caption{\label{fig:22440}(color online). Densities and correlation potentials at time $t=22.440$au. The top-left shows the
          scaled density response $\delta n_{\alpha}(x,t)/\alpha$ for the three values of $\alpha$ indicated. The inset zooms in on the
          scaled density response in the outer region. The top-right shows the correlation potentials at
          different field strengths. The bottom-left panel plots
          the scaled correlation potential response, $\delta v_{c,\alpha}(x,t)/\alpha$. The steps of the correlation potentials occur
          where the density response is non-linear. The lower right panel
          plots the deviation from linearity, $M_\alpha$, of Eq.~\ref{eq:measure}.}
\end{figure}
We choose a weak field strength $\epsilon_1 = 0.0067$au such that the
predominant response of the system is linear. We then apply weaker
fields, $\epsilon_\alpha$, of strengths: $\epsilon_{0.50} = \frac{\epsilon_1}{2}$ and
$\epsilon_{0.25} = \frac{\epsilon_1}{4}$. 

The top left panel of Figure~\ref{fig:22440} shows that the density
response, defined as $\delta n_\alpha(x,t) = n_\alpha(x,t) - n(x,0)$,
predominantly scales linearly with the field strength: plots of
$\delta n_\alpha/\alpha$ lie essentially on top of each other.  The
correlation potential response, in the lower left panel, in region
$\approx (-5,5)$ also scales linearly with the applied field but deviates
from linearity outside this region, displaying step and peak
structures; these are also evident in the full correlation potential
plotted in the top right panel.  Zooming into the tail regions of the
densities (see e.g. inset of top panel), we see in fact the density
response is not linear in these regions. The steps and peaks in the
non-linear region do not scale with the field strength; we do not
expect them to, as the response is not linear, and they also do not
have any higher-order consistent scaling behavior with the field
strength.

We have checked that the step features are not numerical artifacts:
they are converged with respect to the size of the box and
grid-spacing. Changing these parameters may change the details of the
noise in the small oscillations visible in $\delta v\c$ (much smaller
scale than the scale of the step itself) but do not change the overall
structure. This is true for all the graphs shown in the paper.

To quantify the deviation from linearity we next define a measure,
which we plot in the lower right panel. Since the weakest strength is
closest to the ideal linear response limit, we define the deviation
relative to this strength, and define: 
\ben 
M_\alpha = \frac{|\delta
  n_{\alpha} - 4\alpha \delta n_{0.25}|}{|\delta n_{\alpha}|+ 4\alpha
  |\delta n_{0.25}|}. 
\label{eq:measure} 
\een 
If the density response at field strength $\alpha$ was truly linear,
the numerator would vanish (within the approximation that when
$\alpha=0.25$ the system response is linear); and it is trivially zero
when $\alpha = 0.25$.  The measure takes values from 0 to 1, growing
as the degree of non-linearity grows. Note that when the signs of
$\delta n_\alpha$ and $\delta n_{0.25}$ are opposite, the measure
takes the value of 1.  In the lower right panel in Figure
~\ref{fig:22440}, we see that, aside from a sharp peak structure near
$x=0$, $M_\alpha$ is small in the region $x \approx (-5,5)$, then grows
outside this region, peaking and remaining large after the peak. The
sharp structure near $x=0$ occurs due to the density responses
themselves going through zero near the origin. The step structures in the correlation potential appear only in the outer region, where the measure is appreciable, i.e. the density response is significantly non-linear.

\begin{figure}[h]
	\begin{center}
	\includegraphics[width=0.475\textwidth,height=0.4\textwidth,clip]{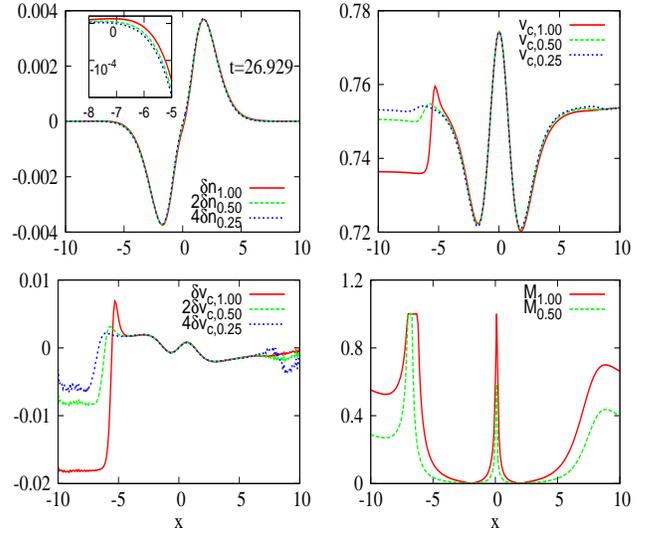} 
	\end{center}
	\caption{\label{fig:26929}(color online). Densities and correlation potentials at time $t=26.929$au. See caption of Fig.~\ref{fig:22440} for details. }
\end{figure}
\begin{figure}[h]
	\begin{center}
	\includegraphics[width=0.475\textwidth,height=0.4\textwidth,clip]{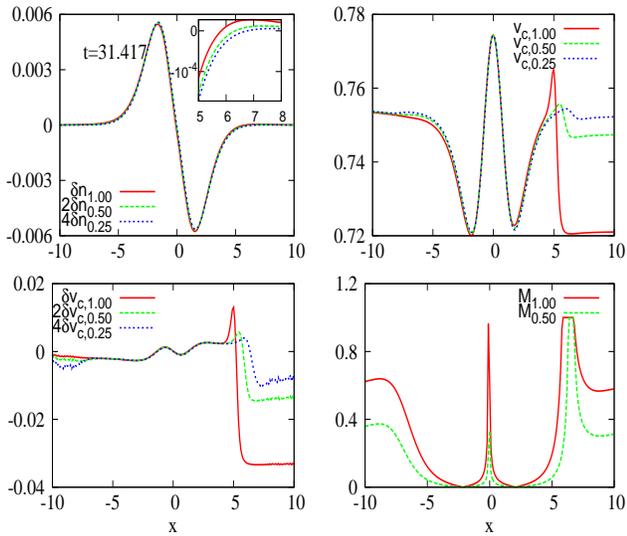} 
	\end{center}
	\caption{\label{fig:31417}(color online). Densities and correlation potentials at time $t=31.417$au. See caption of Fig.~\ref{fig:31417} for details.}
\end{figure}

Figures~\ref{fig:26929} and~\ref{fig:31417} show the density responses
and correlation potentials plotted in the same way, at two different
times, $t=26.929$au and $t=31.417$au. The same conclusions can be
drawn as for the earlier time, and in fact for all the different times
throughout the time propagation that we analyzed: {\it step structures
  appear only in regions where the system's response is nonlinear}. We
did not find a single time at which steps occurred in a region where
the density response is linear.  The step structures do not scale in
any consistent way with the field strength.  (Where the system
response is linear, the correlation potential response scales linearly
with the field, as expected).  There are times at which the step is
abnormally large: this tends to happen in close-to-nodal structures of
the density, and is likely a feature only of two-electron systems.

We note that regions of non-linear system response are typical in
linear response calculations: essentially, the term representing the
field in Hamiltonian $H_0 + {\cal E}(t)x$ gets larger than the
field-free term for large $x$, so a perturbative treatment of it in
that region is no longer valid. However, such a calculation is still
considered to be in the linear response regime, since these regions
contribute negligibly to practical observables extracted from the
system dynamics.

\subsection{Dynamics under a ``delta-kick''}
\label{sec:delta}
\begin{figure}[h]
 \begin{center}
  \includegraphics[width=0.3\textwidth,clip]{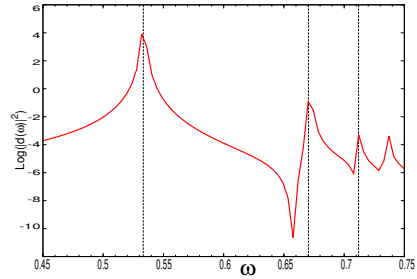} 
 \end{center}
 \caption{\label{fig:spec}(color online). The dipole power spectrum obtained from solving the time-dependent 
 Schr\"odinger equation. Vertical dashed lines indicate the dipole-allowed singlet transition energies, which agree with the energy spectrum.
 (Note the relative oscillator strengths are not accurate because the propagation time was not long enough.)}
\end{figure}

A common way to obtain linear response spectra from real-time dynamics
is to apply a ``delta-kick'' to the system at the initial time, and
measure the subsequent free evolution~\cite{YNIB96}.  That is, ${\cal
  E}(t) = k\delta(t)$, so that we can write $\Psi(t=0^+) =
e^{ik\hat{x}}\Psi(t=0)$. For small enough kick strengths $k$, the system
response is linear in $k$.  Fourier transforming the time-dependent
dipole moment yields the spectrum shown in Fig~\ref{fig:spec}, where a value of $k=0.01$ was
used.  The peaks correspond to the singlet excited states of
odd parity as these are dipole-allowed. The peak-frequencies shown can be confidently assigned to these states only up to about $\omega \sim 0.73$au, because the excited states of energies higher than this have spatial extent too large for the size of the box in our calculation (we have checked convergence with respect to box size for the lower excitations).
Now we consider the same analysis as in the previous case: we halve
$k$ and study the response of the correlation potential and density,
looking for the step feature. The main difference from the Gaussian
pulse field is that now all the dipole-allowed singlet excited states
are equally stimulated: the power spectrum for the delta-kick is uniform.

\begin{figure}[h]
 \begin{center}
  \includegraphics[width=0.475\textwidth,height=0.4\textwidth,clip]{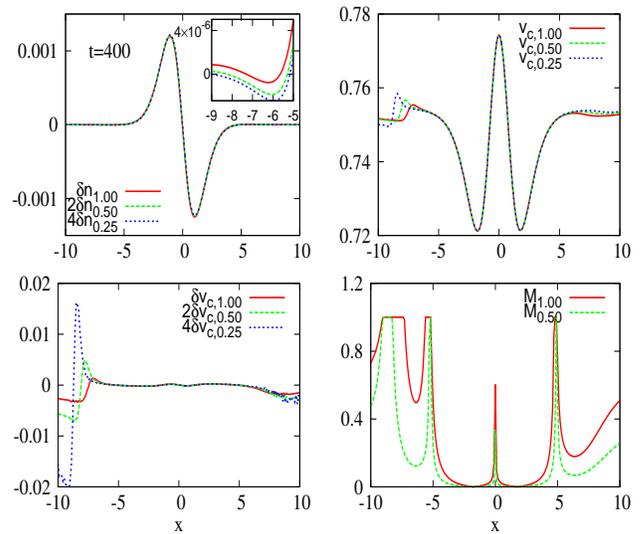} 
 \end{center}
 \caption{\label{fig:kickt1}(color online). At time $400$au after the kick is applied, the response densities and correlation potentials are shown; please refer to Fig.1 for the details of the panels. }
\end{figure}

\begin{figure}[h]
 \begin{center}
  \includegraphics[width=0.475\textwidth,height=0.4\textwidth,clip]{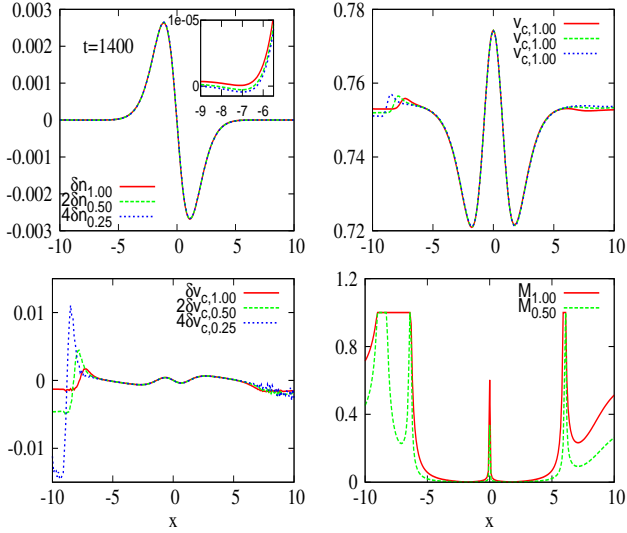} 
 \end{center}
 \caption{\label{fig:kickt2}(color online). As in Figure~\ref{fig:kickt1} but for time 1400 au.}
\end{figure}

Figures~\ref{fig:kickt1} and~\ref{fig:kickt2} show the response
densities and correlation potentials at two snapshots of time 400au
and 1400au, respectively. Similar graphs appear at the other times we
looked at. We again see steps and (sometimes large and oscillatory)
peak-like structures, but, again, they appear only in the region of
non-linear density-response; regions that contribute negligibly to the linear response observables. Once again, these structures are fully
non-linear, in that there is no consistent scaling of their size with
the field strength.


\section{Linear terms in $v\c$ in field-free evolution of a perturbed ground-state}
\label{sec:analysis}
The dynamical step that was found in typical non-linear dynamics situations arises from the 
the fourth term of
Eq.~\ref{eqn:vsexact}, as discussed in Ref.~\cite{EFRM12}. Here we
analyse that term, as well as the full correlation potential, in a linear response situation, by explicitly finding the terms that scale linearly with the deviation from the ground-state.

We consider field-free evolution of a perturbed ground
state, for example, as would occur in the delta-kicked propagation of
the previous section.
We can then expand the wavefunction at time $t$ in terms of the eigenstates,  $\Psi_m$, of the unperturbed system, as 
\ben
\Psi(t) = e^{-iE_0t}\left(\Psi_0 + \sum_{m}c_me^{-i\omega_mt}\Psi_m\right)
\label{eq:psi_exp}
\een
where $\Psi_0$ is the ground-state, $\omega_m = E_m-E_0$ are excitation frequencies,  and $c_m$ are expansion
coefficients, to be considered the small parameter. For example,
in the delta-kick of the previous section, $c_m = ik\langle
\Psi_0\vert\hat{x}\vert\Psi_m\rangle$ (where, for the two electron case
$\hat{x} = x_1 +x_2$). (Note that in the general case, $c_0$ need not be zero).
Then we may write, to first order in the $c_m$, 
\ben
n(x,t) = n_0(x) -2i\sum_{m}c_m\sin(\omega_mt)n_{0m}(x)
\een
where $n_0(x)$ is the ground-state density and $n_{0m}(x) = 2 \int dx'\Psi_0(x,x')\Psi_m(x,x')$ is the $m$th transition density.
Also, we have, to linear order in $c_m$, 
\ben
j(x,t) = i\sum_{m}c_m \cos(\omega_m t) j_{0m}, 
\een
where  $j_{0m}(x) = 2\int dx' \left[\Psi_m \p\Psi_0/\p x -\Psi_0 \p\Psi_m/\p x \right]$.
So, to linear order in the $c_m$, 
\ben
\int^x \partial_t u(x',t) dx'= -i\sum_{m\neq 0} c_m \omega_m\sin(\omega_m t) \int^x \frac{j_{0m}(x')}{n_0(x')} dx'.
\een
If there is any step in the correlation potential that appears at
linear order, it must appear in this term.  From computing just the
excited state wavefunctions and their energies, the right hand side
can easily be computed.  
Further, expanding all terms in Eq.~(\ref{eqn:vsexact}) to linear order, and using Eq.~(\ref{eqn:vcdef}), we get the response of the correlation potential to first order as:
\begin{widetext}
\ben
\delta v\c = \sum_{m\neq 0} i c_m \sin(\omega_m t)
\left(\frac{(\p_x n_0)^2}{2n_0^2}\left(\frac{\p_xn_{0m}}{\p_x n_0} - \frac{n_{0m}}{n_0}\right) 
-\frac{\p_x^2n_0}{2n_0}\left( \frac{\p_x^2n_{0m}}{\p_x^2n_0}- \frac{n_{0m}}{n_0} \right) 
+\omega_m \int^x \frac{j_{0m}(x')}{n_0(x')}dx' + \int \frac{n_{0m}(x')}{\sqrt{(x-x')^2+1}} dx' \right).
\een
\label{eq:vc_exp}
\end{widetext}
(Note that the $c_m$ are pure imaginary, and the correlation potential is indeed purely real). 

\begin{figure}[h]
 \begin{center}
  \includegraphics[width=0.35\textwidth,clip]{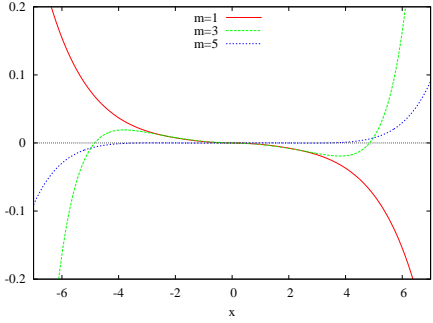} 
 \end{center}
 \caption{\label{fig:sftc}(color online). The correlation responses from first $3$ terms are plotted. }
\end{figure}

Plotting these terms for the delta-kicked soft-Coulomb well, where
$c_m = 2ikd_{0m}$, there is no step seen; as one moves out to larger x
the terms can grow very large, but there is no
step-structure. Figure~\ref{fig:sftc} plots the response correlation potential 
arising from the lowest three dipole-accessible states (which are the first, third, and fifth excitations) in the sum of Eq.~\ref{eq:vc_exp}; the
contributions from higher order terms decrease rapidly, due to the
decreasing oscillator strength.  
Moreover, carrying out the expansion to second-order in $k$ there is also no evidence of step-like structure.
This is consistent with results of previous section; the
regions where there is a step are in fact where such an expansion does not hold, and
the response of the system is fully non-linear.

The results so far show that the dynamical step feature does not
appear in linear response.  That is,  the lack of the non-adiabatic step feature in approximations
does not affect the success of the approximations in predicting linear
response, because this feature only appears in situations where the system response is non-linear. This conclusion has been based on the
model 1D He atom, and we expect it to go through for the general
three-dimensional $N$-electron case. A question might arise about
systems that have states of multiple-excitation character in their
linear response spectra: it is known that for TDDFT to capture such
states the exchange-correlation kernel must have a
frequency-dependence~\cite{MZCB04}, indicating the underlying linear
response exchange-correlation potential has an essentially
non-adiabatic character. For the He atom (1D or 3D), such states
however lie in the continuum and, although they can be accessed by the
delta-kick perturbation~\cite{TK09}, they contribute much less to the spectrum
than the bound states and are outside the range of frequencies for
which our dynamical simulations can be trusted. A better model to
explore states of multiple-excitation character is a 1D model of a
quantum dot: the Hooke's atom, where two soft-coulomb interacting
fermions live in a harmonic potential. The lowest singlet excitation is
predominantly a single-excitation (excitation of the electronic center of mass
coordinate), but the 2nd and 3rd excitations are (largely) mixtures of one
single-excitation and one double-excitation~\cite{EM11,MZCB04}; one is
the second excitation of the center of mass coordinate while the other
is an excitation in the electronic relative coordinate.  A dipole pertubation
applied to such a system can only couple to the lowest excitation in
linear response, a result that can be interpreted in terms of the
harmonic potential theorem~\cite{D94}. A quadratic kick however does
excite the 2nd and 3rd excitations, and this is what we will consider
now: we take
\ben
V(x,t) = \sum_{i=1}^2 \frac{1}{2}\left(1+k\delta(t)\right)x_i^2 
\een
so that in Eq.\ref{eq:psi_exp}, $c_m= i k\langle\Psi_0 \vert{\hat x}^2\vert\Psi_m\rangle$. 
\begin{figure}[h]
 \begin{center}
  \includegraphics[width=0.35\textwidth,clip]{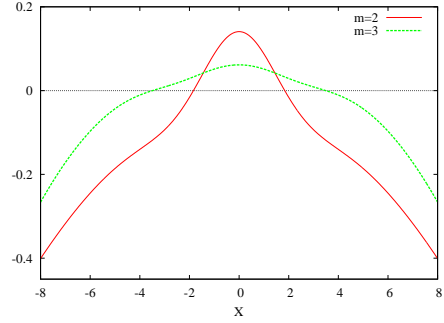} 
 \end{center}
 \caption{\label{fig:detc}(color online). The correlation potential responses from double excitation contributions are plotted. }
\end{figure}
In Figure~\ref{fig:detc}, we plot the contribution to the first-order correlation potential of Eq.\ref{eq:vc_exp} 
of the two states of double-excitation character mentioned above. Once again, there is no step structure 
evident. The non-adiabaticity required to capture states of double-excitation  in linear response is unrelated 
to the dynamical step feature uncovered in Ref.~\cite{EFRM12}.

\section{Summary}
\label{sec:summary}
In this work, we studied the correlation potential of model
two-electron systems in the linear response regime to investigate the role of the
dynamical step feature found in recent studies of
time-dynamics~\cite{EFRM12,FERM13}. We applied a weak field
to the soft-Coulomb helium atom for which we
could extract the exact correlation potential. We found that
step features in $v\c$ only appear in regions far from the
system, in the tails of the density, where the response of the system is in fact non-linear.  These
regions, by definition, do not contribute to  the measured linear response of observables.  These results therefore explicitly justify the
expectation expressed in Ref.~\cite{EFRM12}, that the non-adiabatic
non-local step feature that was generically found there in the
time-dependent correlation potential is a feature of non-linear
dynamics, and is related to having appreciable population in excited
states.  

This explains why adiabatic approximations can usefully predict linear
response spectra in general, while these same approximations may give
incorrect time-dynamics in the non-perturbative
regime~\cite{EFRM12,FERM13,FHTR11,RG12}. The incorrect dynamics observed in
the non-linear regime was due in part to the non-adiabatic non-local
dynamical step found recently in these works, while the results here
show that these are absent in the linear response regime.  States of
multiple-excitation character require a non-adiabatic approximation,
but our analysis of the Hooke's quantum dot model here has shown that
this is unrelated to the appearance of the dynamical step: even in
dynamics where double-excitations appear, the step still is absent in
the linear response region.

{\it Acknowledgments} Financial support from the National Science
Foundation CHE-1152784 (for K.L.), Department of Energy, Office of
Basic Energy Sciences, Division of Chemical Sciences, Geosciences and
Biosciences under Award DE-SC0008623 (N.T.M.), the European
Communities FP7 through the CRONOS project Grant No. 280879 (PE), and
a grant of computer time from the CUNY High Performance Computing
Center under NSF Grants CNS-0855217 and CNS-0958379, are gratefully
acknowledged.

\end{document}